# Simulation of Wheelchair Movements in Crowd Using Fine Grid Cellular Automata

**Dr. Siamak Sarmady**
Department of IT and Computer Engineering
Urmia University of Tech., Urmia, Iran
Email: siamaksarmady@uut.ac.ir

**Associate Professor Dr. Fazilah Haron**
College of the Computer Science and Engineering
Taibah University, Madinah, Saudi Arabia
Email: fabdulhamid@taibahu.edu.sa

**Professor Dr. Abdullah Zawawi Talib**
School of Computer Science
Universiti Sains Malaysia, Penang, Malaysia, 11800
Email: azht@usm.my

## ABSTRACT

Crowd simulation models are used to assess the performance and safety of crowd systems. In some systems, wheelchairs and other moving objects are present in the crowd. The different size and speed of the wheelchairs could significantly change the behavior and dynamics of the crowd. In order to minimize the risks of overcrowding and other types of accidents, it is important to properly model the wheelchairs and their interactions with pedestrians and the environment. Cellular automata are extensively utilized in crowd modeling because of their simple and fast algorithms. Fine grid cellular automata model uses small cells in which moving entities (pedestrians, wheelchairs, cars and etc.) occupy several cells. The entities could have different sizes, shapes, and speeds. In this article, fine grid cellular automata model has been modified to allow building crowd simulation models with different ratios of wheelchairs that could be of different sizes and speed profiles. A scenario of a walkway has been used to evaluate the model. The slow down effect of the slower wheelchairs has been properly reproduced in the results which also match empirical data. Density-speed graphs are also compared to crowds comprising of only pedestrians.

## 1 Introduction

Crowd simulations are used to determine the efficiency of crowd movements in buildings and to predict issues that might arise. These models are also used to estimate the evacuation time of such facilities. In some scenarios and places, different types of wheelchairs could exist in the crowd. The size and the speed profile of wheelchairs are usually different from the pedestrians. They therefore can affect the dynamics of the crowd. The bigger and slower wheelchairs could slow down the crowd and reduce the performance of the crowd system. In a dense crowd, wheelchairs could act as barriers which stop the crowd and create the risk of falling and trampling. Pedestrians could also push and topple wheelchairs and again create the aforementioned risk. In scenarios where significant numbers of wheelchairs might be present in the crowd, it is therefore, essential to model their interactions with the crowd and to ensure a smooth crowd movement. These scenarios could include both normal and evacuation situations. In order to model wheelchairs, a suitable base model that allows the integration of the wheelchairs into the crowd should be used.

Models that produce continuous movements, such as force-based models [1-3], generate smooth movements and could potentially provide accurate and realistic results. These models usually use differential equations. Improvement of these models requires the modification of the relatively complicated equations. In addition, it is computationally expensive to solve systems of differential equations for large number of pedestrians.

Cellular automata models are simpler and more flexible. They use simple transition rules to determine the next cell selected by each agent for its next step. Conventional cellular automata crowd simulation models [4-8] divide the area into a grid of cells of the size of a single pedestrian. The uniform size of the grid cells limits the size of agents and the speed values they might take. Furthermore, the chess like movements are neither smooth nor realistic. In the real world, pedestrians have various sizes, and their average movement speeds could be different. Fine grid cellular automata model still uses the simple movement rules while improving the accuracy. The adaptive speed method adjusts the model to match a given empirical speed-density graph [9].

The multi-layer simulation model of this article uses discrete event methods to simulate the actions of the agents. The finer grained cells and entities that occupy more than one cell allow introducing different types and sizes of moving objects into the crowd such as pedestrians, wheelchairs, shopping trolleys, proms, and cars which have different speeds in different situations. Different body maps (i.e. the cells occupied by an agent) have been prepared for pedestrians and wheelchairs. The simulation results could provide both visual and numerical information about the behavior of the moving entities (i.e. pedestrians and wheelchairs) and the overall details of the crowd system.

The organization of the paper is as follows. Section 2 reviews the existing related work. Section 3 introduces the fine grid model used for wheelchairs. Section 4 presents the simulation results and evaluation of the model. Section 5 concludes the paper.

## 2 Related Work

Crowd simulation models are generally categorized into macroscopic, mesoscopic, and microscopic categories. Macroscopic models consider a crowd as a single liquid-like entity where different parts have varying densities and speeds. Hughes' continuum theory of pedestrian flow is an example of macroscopic models [10]. These models require limited amount of computation to calculate properties of the crowd like density, speed and flow from differential equations [10-15]. Microscopic models on the other hand predict the emerging behaviors of a crowd by simulating the movements and actions of individual pedestrians. The interactions of pedestrians with building structures and obstacles in different situations (e.g. evacuation and normal scenarios) can be observed and studied in these models. It is also possible to estimate macroscopic properties of the crowd (e.g. density, speed and flow) using these models. These models require more calculations but the simulations can still be run on normal desktop PCs.

Mesoscopic models stand somewhere between the above categories. They do not simulate the movements of individual pedestrians, but they consider them in the calculations of flow in different parts of a network comprising of rooms (nodes) and corridors(links). These models use queueing networks and discrete event simulation methods to calculate flow and determine whether congestion and queueing will happen at corridors and doorways. Lovas [16] built a model that simulates the way selection of pedestrians and uses the information to calculate the flow of each path. Hanicsh et al. [17] provide further information such as the average density, average speed and average flow of each region in their model.

More work has been done on microscopic models compared to the other categories. There are a few major approaches in this category which include physics based, cellular automata and rule-based models. Physics-based models use physical rules to predict the movements and interactions of individual pedestrians and the emerging behaviors of the crowd. Social forces [1], magnetic force [2], forces method [3] and velocity obstacle model [18-22] are among these methods. Helbing et al. [1] use imaginary attraction and repelling forces which are applied to pedestrians to simulate their movements. These forces include the repelling force between pedestrians, repelling forces between pedestrians and obstacles (collision avoidance), and the attraction force between a pedestrian and its goal.

The second approach in the microscopic category includes models that divide the space into cells and utilize cellular automata rules [4-8], distance maps [23-25], or similar methods to simulate the movements of pedestrians among the cells. There are three major type of these models. In the first type of cell-based models, only a single pedestrian could be accommodated in each cell. There is no need for collision avoidance methods since it is automatically avoided. The second type of cellular automata models allow more than one pedestrian to occupy a cell [26, 27]. These models are similar to mesoscopic models and as such, provide less information about the movements of individual pedestrians. In the last type of cell-based models, each pedestrian occupies several cells. Kirchner et al. [28] proposed such a model in which pedestrians would occupy an identical 2 x 2 cell area. They concluded that the model is more accurate and it falls in the gap between regular cellular automata models and continuous models. The transition rules in cellular automata models are a set of simple mathematical equations or logical rules that determine the next cell for a pedestrian. Since these rules and equations are evaluated independently for each pedestrian, these methods could be very fast. Some of the models that used cellular automata to simulate crowd and vehicle movements include those by Nagel[29], Biham [30], Nagatani [31], Dijkstra [6], Blue [4], Kirchner et al.[5, 7, 8]. A major difference between these models, is the cell transition algorithm. A popular approach is to calculate the distance of each of the neighboring cells to a desired target and select the neighbor cell nearest to the target every time or with a higher probability. Sometimes a map that determines the distance of each cell in the movement space to a target is precalculated. Kretz et al. [25] have reviewed the different methods used for creating distance maps.

Models in the last microscopic crowd simulation approach utilize certain behavioral rules to simulate flocks of birds, groups of fishes, or herds of animals. An example of these methods, Reynolds' rule-based model uses rules that include collision avoidance, velocity matching with other flock members and flock centering (i.e. staying near other members) [32]. However, these models are mostly used to make animations. They can hardly produce realistic simulations of collision avoidance, density effects and behaviors of pedestrians in narrow passages.

The cellular automata models reviewed in this section do not support a mix of pedestrians with moving objects like wheelchair. In this research, cellular automata model is extended to support these features. The model can therefore allow studying new situations and provide more realistic results.

Shimada et al. [33] have performed an experimental study about the effect of wheelchairs on the crowd movements. Crowds of pedestrian and pedestrian-wheelchair mixes with different mixture ratios and different densities have passed from corridors and gates with different width in the study. In addition, two different types of wheelchair (i.e. assisted and non-assisted manual) have been used in the experiments. The flow of crowd is measured in every case and compared. The data in this article is used for the validation of the proposed wheelchair-pedestrian simulation models in this research.

## 3 Models

The main goal of this research is to extend the fine grid cellular automata model[34, 35] to allow simulation of a crowd comprising of pedestrians and wheelchairs. A high-level multi-layer model of the actions and movements has been used for moving entities. Discrete event methods are used to simulate actions of entities (i.e. pedestrians, wheelchairs and etc.) while the movements have been simulated using fine grid model.

### 3.1 Fine Grid Model for Wheelchairs

The smaller cells in fine grid model allow more speed levels (i.e. cell transitions per second) and therefore better adjustment of speed. A cell size of 5 cm × 5 cm and a 0.025 s time step have been used in this study. There are therefore 41 possible speed values (0 - 40 cells per second). Different profiles can be created for pedestrians and other moving entities representing different classes of them (e.g., children, teenager, and adult profiles for pedestrians) with varying body shapes, sizes, and speeds.

For each moving entity, several maps could be created for moving towards different directions. In this research, only eight maps for eight major moving directions have been created. There are currently three moving entity types which include pedestrians, non-assisted manual or motorized wheelchairs, and assisted wheelchairs. Figure 1 shows the maps used for the two wheelchair types. These maps are used in the simulations depending on the movement direction of entities. In each time step, the movement direction is determined. Then the center point of each entity's map is moved to one of its eight Moore neighboring cells. Notice that there are different type and sizes of wheelchairs and each type might have different average free flow speed. Motorized wheelchairs naturally have higher free flow speed than normal pedestrians while manual non-assisted wheelchairs are usually slower. The cellular automata transition rules are exactly the same as pedestrians, except that wheelchairs have different profiles (e.g. body maps and average speed). It is possible to create different mixtures with varying ratios of different profile types to represent the crowd in different places and scenarios.

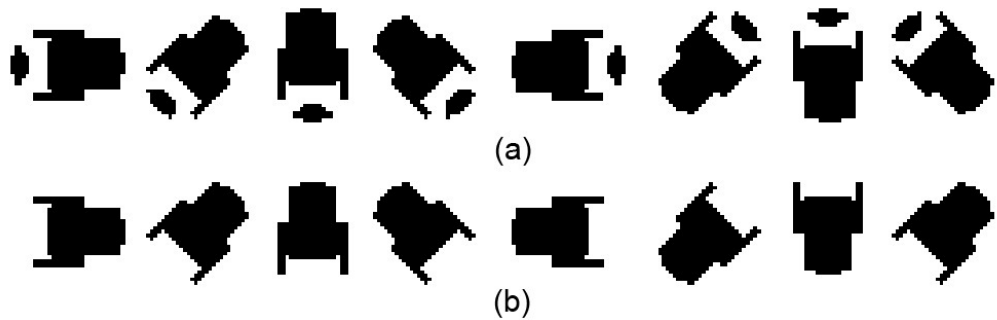

Figure 1: Maps for assisted manual (a) and non-assisted or motorized (b) wheelchairs on a 5cm grid

In fine grid model, an empirical density-speed graph is chosen (e.g. fruin's) and the density-speed relation of pedestrians is adjusted to match the graph. At each moment, the density perceived by each pedestrian is measured in the simulation, and a corresponding speed is drawn from the selected graph. Then the movement speed of each pedestrian is adjusted to the speed determined by the graph. A similar method is used for wheelchairs. However, the movement speed of wheelchairs in different perceived densities and their free flow speed (i.e. density=0) are not the same as pedestrians.

According to [36], manual wheelchairs (self and assistant propelled) have an upper speed limit of 6km/h (1.66 m/s) and powered wheelchairs may have an upper speed limit of 12km/h (3.33 m/s) in the UK. These limits are different in other countries. In addition to the limits, the maximum speeds of powered wheelchairs are different from one model to the other (Figure 2). Racing wheelchairs have even higher maximum speeds which could exceed 8.5 m/s [37]. However, the free flow speed of wheelchairs could be much lower than their maximum speed. The average and the free flow speed of wheelchairs depend on the location and the movement purpose. One experiment measured the speed of wheelchairs while crossing a road with a 10 meters width to be 1.19 m/s for motorized and 1.083 m/s for self-propelled wheelchairs [38]. Another study estimated the average speed of self-propelled wheelchairs to be around 0.5–0.8 m/s [39].

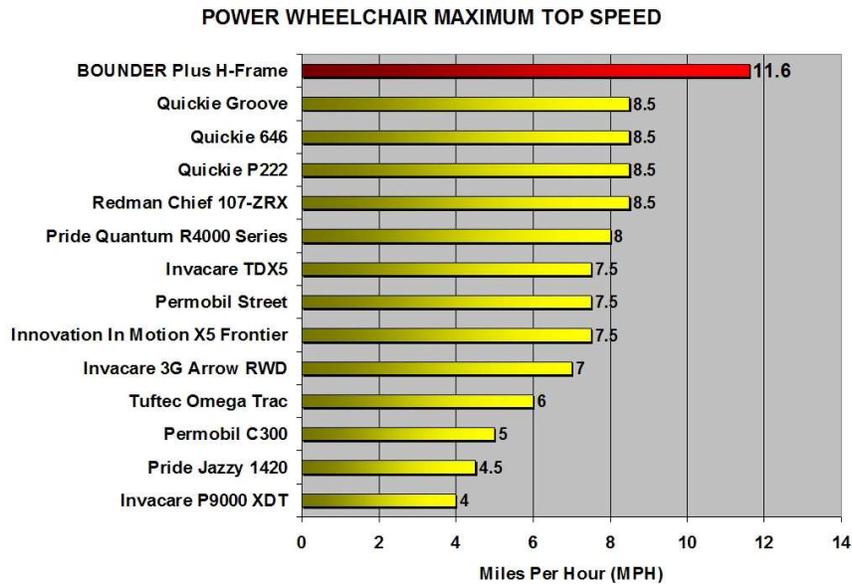

Figure 2: Maximum speed for different motorized wheelchair models [40]

The fine grid model uses empirical density-speed relations like Fruin's diagram to adjust the movement speed of pedestrians. A density-speed diagram is therefore needed to determine the speed with which the wheelchairs will move for different precepted densities. Such graphs are not available for the time being, so the approach taken in this research is to scale the empirical graph that is chosen for pedestrians in a simulation (e.g. Fruin's) for the wheelchairs. If the majority of moving entities in a crowd are pedestrians, then it seems natural to assume that the density-speed diagram of wheelchairs has almost the same shape as the diagram for pedestrians (except that it is scaled with the free flow speed of wheelchairs).

Firstly, the free flow speed of wheelchairs (i.e. unimpeded speed at zero precepted density) is determined. This will be the starting point of the density-speed graph for wheelchairs (i.e. the intersection with vertical axis). If the majority of the moving entities in a crowd are pedestrians, then the lower extreme of the wheelchair density-speed graph (i.e. stall density or the intersection with horizontal axis) will be determined by pedestrians. That is because when the crowd stops the wheelchairs will not be able to move.

Let's assume Fruin's graph is selected for pedestrian movements in a fine grid simulation scenario. Also let's consider that, two type of wheelchairs with free flow speed of $V_1$=1.5 m/s and $V_2$= 0.8 m/s will be used in the simulation. One of these types has a free flow speed that is higher than the free flow speed of pedestrians (in Fruin's graph) while the other has a lower speed than pedestrians. These values will determine the initial point of the density-speed graph for each wheelchair type. Furthermore, in Fruin's graph, when the density reaches 4 pedestrians/m², the crowd movement stalls. So, the intersections of the graphs for the two wheelchair types with the vertical and horizontal axis of the density-speed diagram are now known.

As described earlier, it is assumed that the shape of the density-speed graph of wheelchairs, is similar to the fundamental pedestrian density-speed graph that is used for a simulation scenario. The Fruin's graph is therefore scaled in a way that the two extremes (i.e. intersections with vertical and horizontal axis) match those determined above. Figure 3 shows the scaled graphs for the two wheelchair types along with the Fruin's graph for pedestrians. The scaling factor is the ratio of the free flow speed of each of the wheelchair types to the free flow speed of pedestrians. The ratio is then multiplied to the values on the Fruin's graph,

$$V_{Wheelchair} = \frac{V_{ff-wheelchair}}{V_{ff-pedestrian}} \; V_{Pedestrian} \quad (1)$$

This method is also used when other empirical density-speed graphs (e.g. Weidmann's graph) are used for the pedestrians in a simulation scenario. Figure 4 shows the resulting graphs for the two wheelchair types in comparison to the Weidmann's for the pedestrian. During the simulation, the speed of each entity is extracted from either the pedestrian density-speed graphs or the scaled versions, depending on the enity profile.

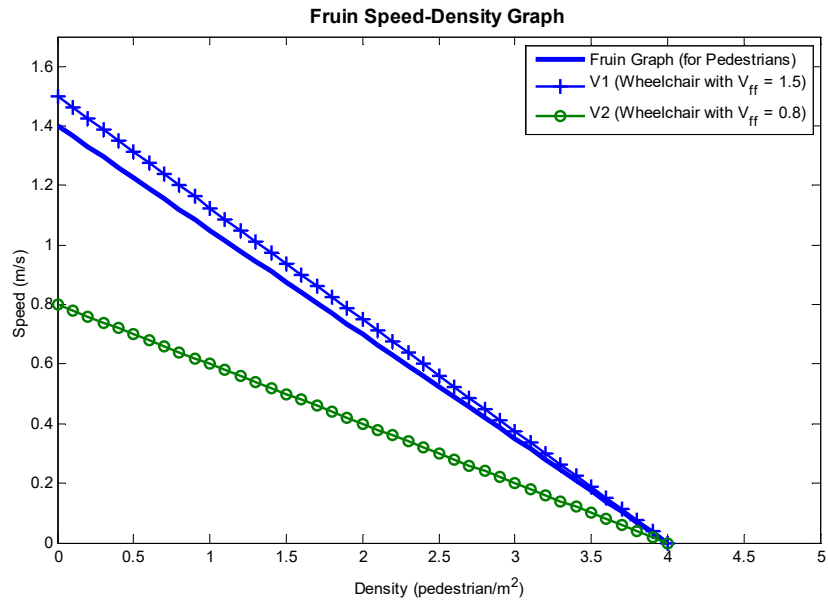

Figure 3: Fruin's graph compared to graphs for wheelchairs with different free flow speeds

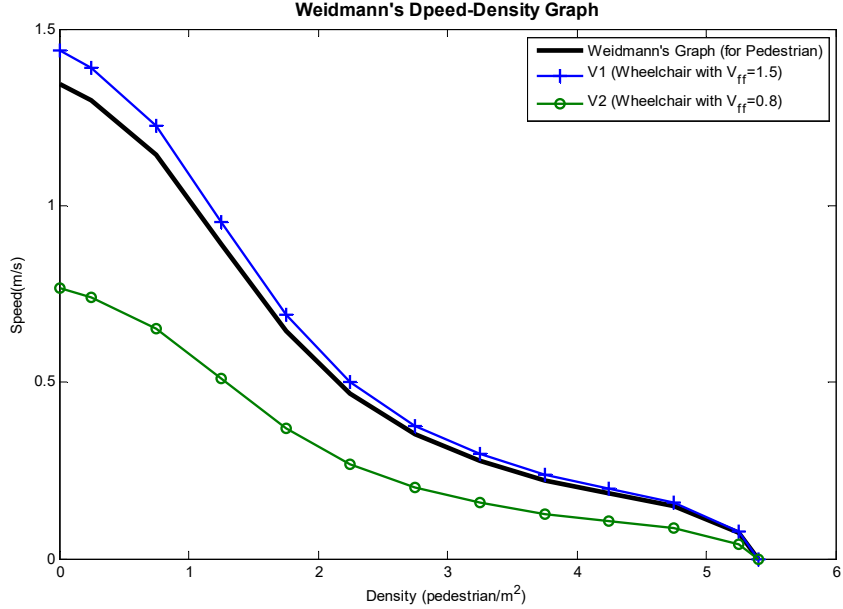

Figure 4: Weidmann's graph compared to graphs for wheelchairs with different speeds

## 3.2 Probability Model for Cellular Automata Transitions

Transition probability $P_i$ is the probability of the center point of the pedestrian map moving into neighbor $i$. It is calculated using the simplest method proposed in the fine grid model. For each neighbor $i$, the parameter $M_i$ is calculated using the following equation [41]:

$$M_i = \frac{n_i}{R_i} \quad (2)$$

$$n_i \in \{0,1\}, \quad R_i \neq 0$$

In equation 2, $R_i$ is the distance between cell $i$ and the target. The parameter $n_i$ is the collision avoidance term, and it is 0 if the neighbor i is occupied and 1 otherwise. The cell nearest to the target gets a larger $M_i$ than the others. In the next step, the neighbors are sorted according to their $M_i$ value in descending order. Given that $M_i$ is directly proportional to the desirability of a cell, the cell with the highest $M_i$ and superior rank (i.e. smaller index) should be selected most of the time. The Poisson statistical distribution (with a $\lambda < 0.1$) is used to randomly select the next cell. A SoftMax function can also be used to calculate the probabilities but it will possibly be more expensive considering that it is called millions of times.

As mentioned in the conditions, the equation 3 is only used when $R_i \neq 0$. $R_i = 0$ means that neighbor $i$ is the target. In that case, cell $i$ gets a probability of 1 if it is not occupied. All other cells get a probability of 0, and the pedestrian moves into the cell which is on the target area.

In conventional cellular automata models, pedestrians will always move with their free flow speed unless they encounter an obstacle or another pedestrian. In order to produce more realistic behavior, fine grid cellular automata, calculates the density of pedestrians and other moving entities in the perception area (towards the movement direction) and uses the available empirical graphs to obtain the speed with which the pedestrian or other moving entities should move. It is then attempted to achieve that speed for the specific moving entity. Further details on the method could be obtained from the original article [34].

## 4 Simulation Results

As described earlier, different types and models of wheelchairs have different average, maximum and free flow speeds. Several studies have been performed on different aspects of wheelchairs but little information is available for the calibration of crowd simulation models involving wheelchairs. Density-speed diagrams for different mixture ratios of pedestrian and wheelchairs could specifically be beneficial for this research. One of the few research articles available is by Shimada et al.[33]. However, the paper lacks the required details for our purpose including the speed specifications of the wheelchairs. Furthermore, density and speed conditions for the experiments and provided graphs are not known. Since no better data was found to be useful, the simulation results are compared to some of the results of this article. In the article, flows of crowd with different pedestrian and wheelchair ratios in two scenarios, namely flow of crowd at a corridor and evacuation of a room with a narrow door are studied.

In the simulation experiments of this research, a walkway or corridor with a width of three meters has been used. Two types of assisted and non-assisted wheelchairs are also used. The free flow speed of assisted and non-assisted wheelchairs are 1.083 and 0.8 m/s, respectively (matching those mentioned in [38] and [39]). It should be noticed that these speeds are used as examples and many other wheelchairs with varying sizes and speeds are used in the reality. A fine grid with cell size of 5cm has been used for the simulations. Figure 1 shows the maps used for the two types of wheelchairs. The Weidmann's density-speed relation has been used to adjust pedestrian speeds and a scaled relation is used for the wheelchairs.

Figure by Shimada et al. [33] shows the relation between the crowd flow and the wheelchair-pedestrian ratio. Two sets of curves have been provided in the figure. Those marked with solid lines are for evacuation scenarios while the dotted lines are for normal crowd flow at corridors with different width. Since we experiment on normal crowd flow and not emergency evacuation, only the second set is relevant to our research. The authors suggest that the width of the corridor does not affect the crowd flow significantly. Furthermore, it can be seen that the flow decreases moderately as the ratio of wheelchairs is increased in the crowd.

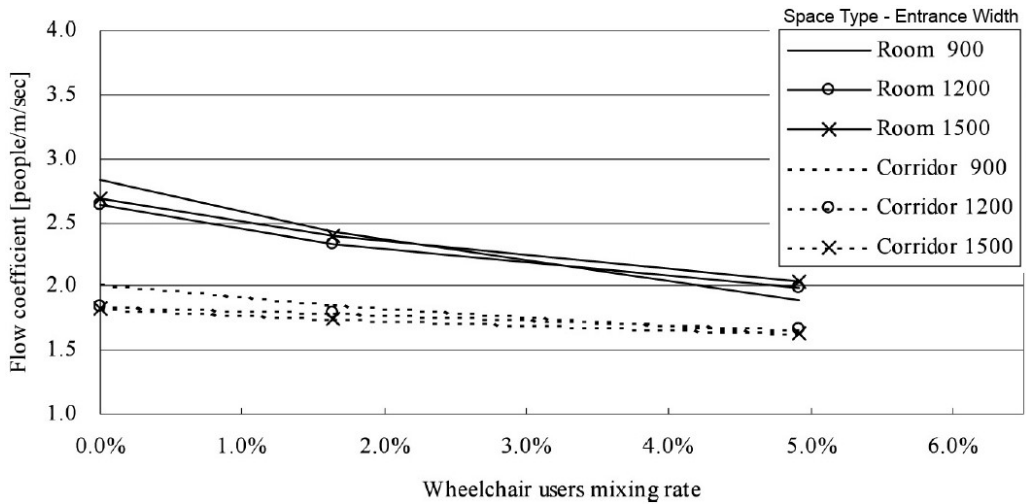

Figure 5: The flow for different mixture ratios of wheelchairs [33]

In Shimada et al.'s work, the maximum amount of flow in corridor scenario is between 1.8-2.0 pedestrians/ms$^{-1}$. These values are much higher than those in the existing empirical fundamental graphs including SPFE[42], Weidmann [43], Predtechenskii and Milinskii (PM) graphs [44], and the data points provided by Older[45] and Helbing [46] (Figure 6). This is perhaps because Shimada et al. tried to find the maximum achievable flow in the corridor case.

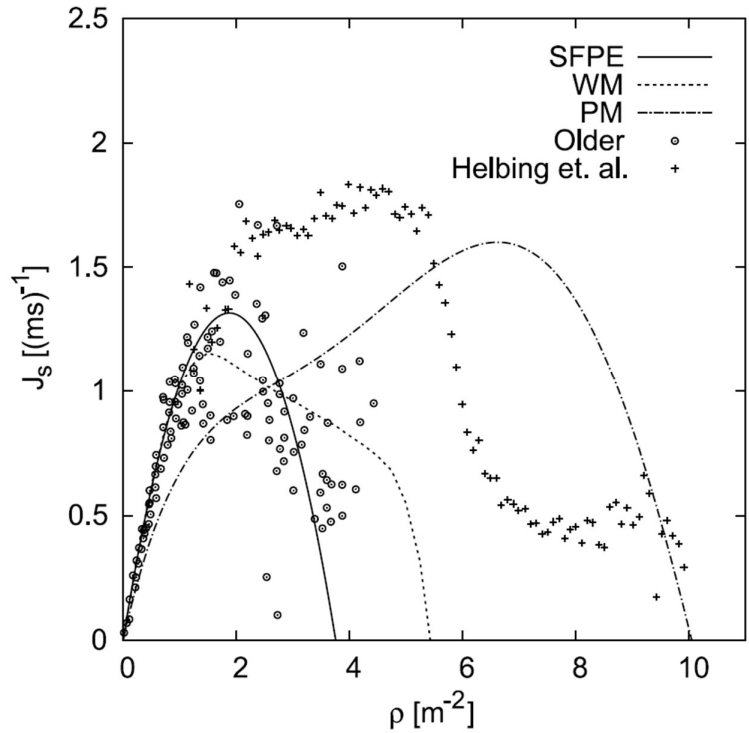

Figure 6: The fundamental graphs in terms of flow-density [47]

A snapshot of the simulation is shown in Figure 7. It can be observed that the pedestrians have been slowed down by the bigger sized, slower moving wheelchairs (notice the jam behind the wheelchairs). Increasing the ratio of wheelchairs will slow down the crowd even further. Larger mixture ratio of wheelchairs increases the slowdown and reduces the average flow of the system. This observation matches the empirical study of Shimada et al. [33] (Figure 5).

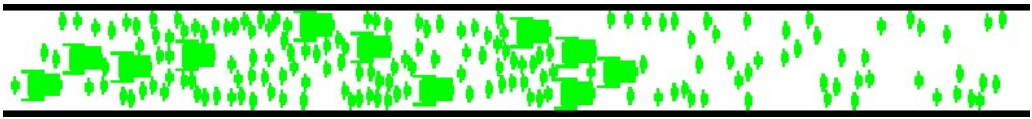

Figure 7: Snapshot of the pedestrian-wheelchair crowd simulation

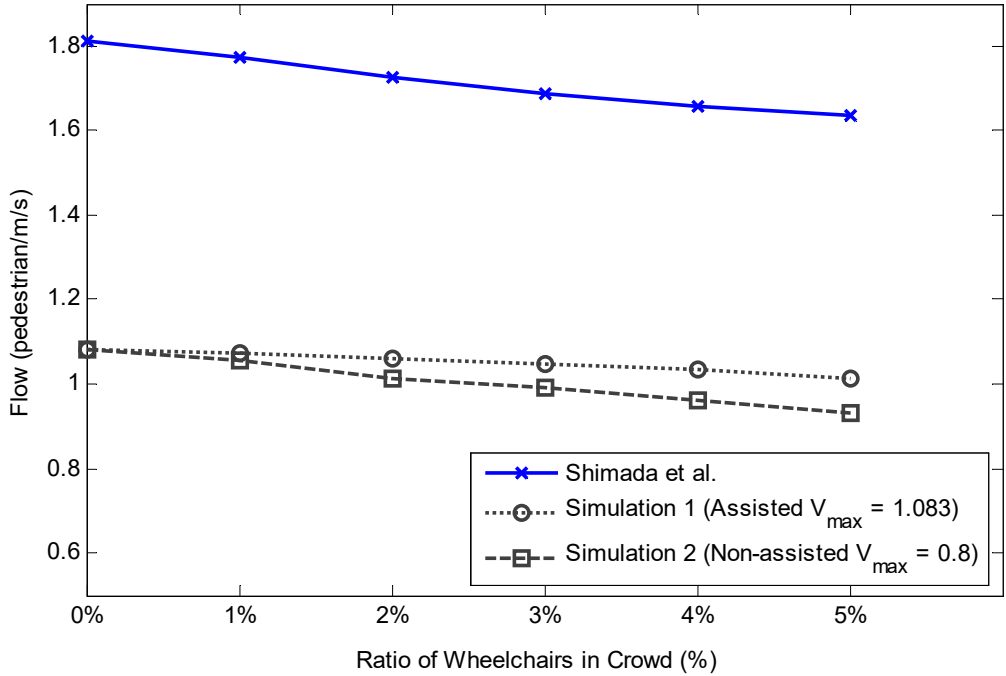

Figure 8: The effect of the ratio of wheelchairs on the crowd flow

The flow of crowd in the simulation is compared to the data provided by Shimada et al. [33] in Figure . For the simulation experiments, it is assumed that the density-speed relation of the crowd that is being simulated matches the Weidmann's data. The maximum flow for a crowd comprising of only pedestrians is around 1.1-1.2 ms$^{-1}$ (Figure 6). This means that around 1.1 pedestrians pass in a second from a 1-meter wide passage.

The flow in the graph is calculated by averaging the flow of crowd during a 1500 seconds simulation. The maximum flow for a 0% wheelchair matches the Weidmann's data. For crowd comprising of pedestrians and different ratios of the two types of the wheelchairs (i.e. assisted and non-assisted), separate graphs are provided.

The slope in the graph of the assisted wheelchair ($V_{max}$= 1.083 m/s) is slightly lower than Shimada's. The moderate slope is possibly because the free flow speed of this type of wheelchair is quite near to the free flow speed of pedestrians (1.344 m/s). For the non-assisted manual wheelchairs which are slower ($V_{max}$= 0.8 m/s), the slope is comparable to Shimada's results and more significant. The slowdown of the crowd due to presence of this type of wheelchair is more considerable. It appears that the proposed simulation model has been able to reproduce the slowdown effect of slower wheelchairs on the crowd.

The density-speed graph of the simulations with the two types of wheelchairs with a mixture ratio of five percent is shown in figure 11. As expected, the general form of the two graphs are similar to the Weidmann's empirical density-speed graph which is used for pedestrians (i.e. the majority of the crowd). Since both wheelchair types are slower than pedestrians, they have reduced the movement speed of the crowd in different densities. The reduction in speed depends on how much slower the wheelchairs are in comparison to pedestrians.

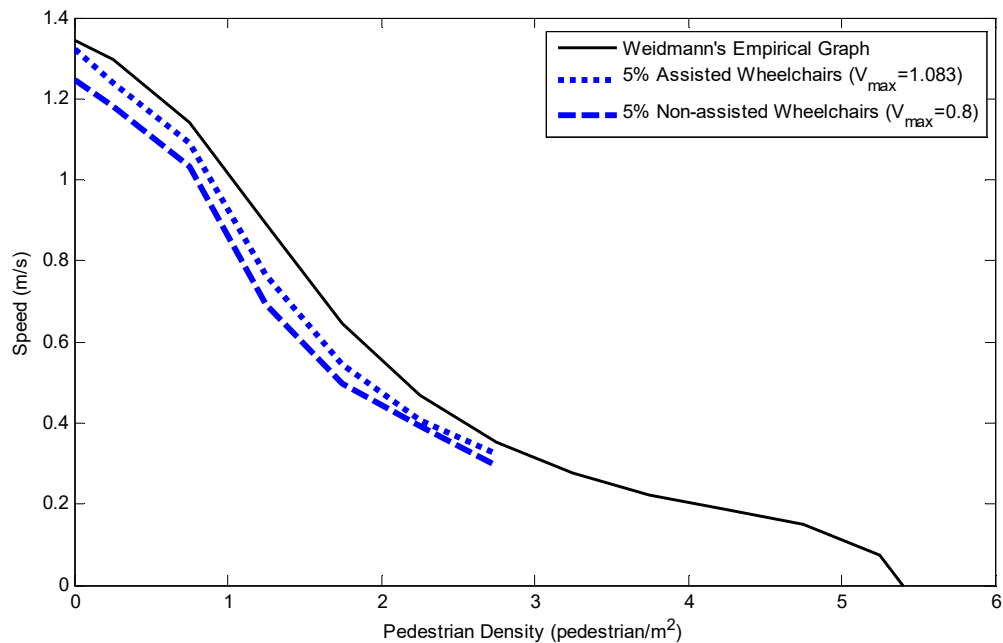

Figure 9: Simulation based on Weidmann's data with 5% wheelchairs

## 5 Conclusion

In this article, fine grid cellular automata model was extended to include other moving objects like wheelchairs, shopping trolleys and push-chairs. The concept of profiles was used to include moving objects with different sizes and speeds in the simulations. The simulation of the movements of the crowd consisting of different ratios of wheelchair and pedestrian were compared to the empirical results. It was shown that the model can reproduce the slowdown effect of the slower-moving wheelchairs on the crowd.